\documentclass[a4paper,11pt]{article}
\pdfoutput=1 

\usepackage{jheppub} 

\usepackage[T1]{fontenc} 

\usepackage{braket} 

\DeclareMathOperator{\scalar}{\varphi}

\title{\boldmath  Proca Theory from the Spinning Worldline}

 \author[a,1]{Matthias Carosi\note{Corresponding author.}}
 \author[b]{and Ivo Sachs}
 \affiliation[a]{Physik Department T70, Technische Universit\"at M\"unchen,\\ James-Franck-Str., D-85748 Garching, Germany}
\affiliation[b]{Arnold-Sommerfeld-Centerfor Theoretical Physics, Ludwig-Maximilians-Universit\"at of Munich,\\Theresienstr. 37, D-80333 M\"unchen, Germany}

\emailAdd{matthias.carosi@tum.de}
\emailAdd{ivo.sachs@physik.lmu.de}

\abstract{We obtain Proca field theory from the quantisation of the $\mathcal{N}=2$ supersymmetric worldline upon supplementing the graded BRST-algebra with an extra multiplet of oscillators. The linearised theory describes the BV-extended spectrum of Proca theory, together with a St\"uckelberg field. When coupling the theory to background fields we derive the Proca equations, arising as consistency conditions in the BRST procedure.  We also explore non-abelian modifications, complexified vector fields as well as coupling to a dilaton field. We propose a cubic action on the space of BRST-operators which reproduces the known Proca action.  }

\begin{document} 
\begin{flushright}
	LMU-ASC 43/21\\
\end{flushright}
\maketitle
\flushbottom

\section{Introduction}
\label{sec:intro}
The quantisation of the relativistic particle  with $\mathcal{N}$-extended worldline supersymmetry is known to describe a particle of spin $\mathcal{N}/2$ propagating in spacetime \cite{Brink:1976uf,Gershun:1979fb}. This has been analysed extensively in many works,  \cite{
Barducci:1976xq,Brink:1976uf,Henneaux:1987cp,PIERRI1990421,doi:10.1142/S0217751X88001132,Bastianelli:1991be,Bastianelli:1992ct,Strassler:1992zr,Reuter:1996zm,Marnelius_1994,Sato:1998sf,Schubert:2001he,Bastianelli:2002fv,Bastianelli:2002qw,Bastianelli:2005vk,Bastianelli:2005uy,Bastianelli:2013pta}  and leads to an alternative description of quantum field theory which is close, in nature, to some formulations of string field theory. In addition, the field equations implied by nilpotency of the worldline BRST charge reproduce the non-linear Yang-Mills equations \cite{Dai_2008} and extensions thereof \cite{Grigoriev:2021bes} for $\mathcal{N}=2$ spinning particle, Einstein equations \cite{Bonezzi_2018} for $\mathcal{N}=4$ spinning particle or the NS-sector of Supergravity equations of motion \cite{Bonezzi:2020jjq}. 

An attractive feature of the BRST-quantisation of the worldline is that it automatically produces the complete spectrum of spacetime fields required for the Batalin-Vilkoviski (BV) quantisation of the corresponding spacetime quantum field theory (\cite{Bonezzi:2020jjq} and references therein).

While not the main focus of this work the BV formalism also gives a very natural way of building interaction terms and of generating a $1$-loop effective action for the theory. Indeed, the worldline approach to computing scattering amplitudes has long been known for its effectiveness \cite{Bastianelli:2013pta,H_lzler_2008,Papadopoulos:1989rg,Schubert:2001he,Strassler:1992zr}.

In the present note we explore a {\it massive}  $\mathcal{N}=2$ supersymmetric worldline theory. 
One way to produce a mass term, is to consider a $1+D$ dimensional spacetime, and set the $D$-th component of the momentum $p_D=m$, analogous to Kaluza-Klein compactification. A similar procedure is followed in ref. \cite{Brink:1976uf,Bastianelli:2005uy}.
While this procedure reproduces the correct BV-spectrum of the Proca theory together with the linearised equations of motion, it  does not allow to promote them to background fields in the BRST operator $Q$ as required for a background independent formulation that produces the non-linear equations of motion by nilpotency of $Q$ as in \cite{Dai_2008,Bonezzi_2018,Bonezzi:2020jjq,Grigoriev:2021bes}. 

In our approach we induce massive fields by enlarging the algebra of the BRST quantisation of the worldline with additional bosonic and fermionic oscillators which can be thought of as describing internal degrees of freedom of the spinning particle. This approach for describing dynamical systems can be viewed as a concrete application of the general framework described in \cite{Grigoriev:2021bes}. Alternatively, one can view it as originating from the BRST quantisation of some extended worldline theory. Either way, analysing the corresponding constraint algebra and the associated BRST charge we obtain in this way a Proca theory in spacetime, whose dynamics is determined by consistency conditions of the BRST quantisation rather than a variation of some Lagrangian. This procedure will prove lengthy, and quite involved at times, but allows us to analyse aspects of quantisation and symmetry which remain otherwise obscure in the Lagrangian formulation of the theory. In particular, the mass of the vector field arises due to the excitation of an "internal" degree of freedom, similar to string theory. Another observation is that a St{\" u}ckelberg field is automatically present in the BV-extended spectrum of the worldline theory and so is the Yang-Mills ghost. The latter is, however, set to zero in the BRST cohomology. 

We then obtain a cubic action on the space of BRST operators which reproduces the known Proca action for a massive vector field. However, due to the lack of a non-degenerate, cyclic trace on the operator algebra the latter is not equivalent to the condition of nilpotency of $Q$, in general.

We begin, in section \ref{sec: YM} with a short review of Yang-Mills theory in the worldline formalism. In section \ref{Proca} we enlarge the worldline algebra of oscillators allowing us to include massive vector fields in addition to massless ones in the BRST spectrum of the theory. In section \ref{sec:vec field} we introduce a generic Lie algebra valued and complex background vector field and analyse the constraint equations implied by nilpotency of $Q$. In section \ref{section: coupling to a scalar} we explore possible couplings to scalar fields in addition to the vector background. In this way we recover the St\"uckelberg decomposition described in section \ref{Proca}. In section \ref{sec:action} we discuss possible action functionals on the space of BRST operators, parametrised by the   background fields introduced in the previous sections. Finally, in section \ref{sec: conclusions} we present the conclusions.

\section{Short review of Yang-Mills theory}\label{sec: YM}
It is well known \cite{Dai_2008} that consistency of the BRST-quantisation of the relativistic particle with $\mathcal{N}=2$ worldline SUSY describes Yang-Mills theory in spacetime. To see this, let us recall the super reparametrisation constraints 
\begin{align}
    q&=\theta^\mu\Pi_\mu\,,\qquad \bar q=\bar\theta^\mu\Pi_\mu\,,\qquad 
    H =  \:-\Pi^2 - 2\kappa \theta^{\mu} \bar{\theta}^{\nu} [\Pi_{\mu}, \Pi_{\nu} ]\,,\qquad \kappa\in \mathbf{R}\,,
\end{align}
where $\kappa$ will be fixed by nilpotency of the BRST-charge below. The complex linear combinations of the worldline spinors, $\theta^\mu=\frac{1}{\sqrt{2}}(\theta_1^\mu+i \theta_2^\mu)$, $\bar \theta^\mu=\frac{1}{\sqrt{2}}(\theta_1^\mu-i \theta_2^\mu)$ satisfy the anti-commutation relations 
\begin{equation}
 \{\theta^\mu,\bar \theta_\nu\}=\delta^\mu_\nu\,,
\end{equation} 
and $\Pi_\mu=p_\mu+A_\mu$ are the canonical momenta for the embedding coordinates, $x^\mu$. The algebra of the constraints does not close, even when the Yang-Mills field is on-shell. Nevertheless, the BRST charge, obtained by combining the constraints with their BRST ghosts (see  e.g. \cite{van_Holten_2004} for a review), 
\begin{equation} \label{QYM}
    Q(A) = c H + \bar{\gamma} q + \gamma \bar{q} + \gamma \bar{\gamma} b \,,
\end{equation}
is nilpotent on-shell,  when restricted to an invariant subspace $V_0$ of the Fock space $V$ generated by $\theta^\mu$ as well as the reparametrisation ghosts $c$ and super ghosts $\gamma$ and $\beta$, accounting for the (super) reparametrisations. The ghost numbers and Grassmann parity assigned to $\left(c,b\right),\left(\gamma,\bar\beta\right),\left(\bar\gamma,\beta\right) $ are  $\left( +1 , -1\right), \left(+1 , -1\right),\left( +1, -1\right)$ and  $\left( 1,1\right),\left(0,0\right),\left(0,0\right)$ respectively, and the non-vanishing canonical (anti-)commutation relations are 
\begin{equation}
    \{c,b\} = 1 \,, \qquad [\gamma,\bar{\beta}] = 1 \,, \qquad [\bar{\gamma},\beta] = -1 \,.
\end{equation}
More precisely, we realise $V$ in terms of wave functions $\Phi(x,\theta,c,\beta,\gamma)$ on which the conjugated momenta act as derivatives, i.e.
\begin{equation}
 p_\mu=-i\partial_{x^\mu}\,, \qquad \bar\theta_a=\partial_{\theta^a}\,, \qquad b=\partial_{c}, \qquad \bar\gamma=-\partial_{\beta},\qquad \bar\beta=-\partial_{\gamma}\,.
\end{equation} 
The subspace $V_0$ is then defined as the eigenspace of 
\begin{align}\label{JYM}
    J & = \theta^{\mu}\bar{\theta}_{\mu} + \gamma\bar{\beta} - \bar{\gamma}\beta - \left( \frac{D}{2} - 1 \right) \\
    & = N_{\theta} + N_{\gamma} + N_{\beta} - \left( \frac{D}{2} - 1 \right) \,,
\end{align}
with charge $\left(2-\frac{D}{2}\right)$. On $V_0$ the BRST-operator $Q(A)$ given by \eqref{QYM} is nilpotent provided $\kappa=1$ and the Yang-Mills equation $[\Pi^\mu,[\Pi_\mu,\Pi_\nu]]=0$ is satisfied. The latter equation is implied by the vanishing of the $[q,H]$ commutation relation. From the spacetime point of view this has the interpretation that gluons can consistently propagate on a background that satisfies the Yang-Mills equations while this is is not the case for generic excitations contained in the complement of $V_0$ in $V$. 

\section{Proca theory}\label{Proca}
The purpose of this section is to explore the modification of this formalism for massive vector fields or Proca fields. We first consider the linearised theory which amounts to  analysing the cohomology of $Q_0$. Then, we discuss the equations implied by nilpotency of $Q$. One might argue that there should be no difference between these two approaches since the Proca theory is linear after all. However, we will see that this is not necessarily the case.

In fact, the linearised theory has been analysed before in \cite{Bastianelli:2005uy}. While our construction differs from the latter by the addition of generators in the BRST algebra inspired by string theory, it will nevertheless  reproduce the spectrum of \cite{Bastianelli:2005uy} with additional St\"uckelberg- and auxiliary fields. As a theory of background fields, however, the two approaches are not equivalent. Indeed, we will see that the additional extra generators are necessary to reproduce the Proca theory.
\subsection{Linearised Proca theory}
In this section we generalise the BRST quantisation of \cite{Dai_2008} described in the last section to include a mass term. A mass term can be added to the constraints by the  Kaluza-Klein approach (e.g.  \cite{Barducci:1976qu,Collins:1976yy,Brink:1976uf,Bastianelli:2005uy}). Here, we use a different construction, taking inspiration from string theory. Concretely, we consider the spinning point particle as a truncation of the RNS open superstring (see ref. \cite{green_schwarz_witten_2012}),  keeping only the first excited modes. More precisely, we extend the phase-space by two canonically conjugated complex bosonic variables $(\alpha^a, \bar{\alpha}_a)$ with $(\alpha^a)^{\dagger} = \bar{\alpha}_a$ and commutation relation $ [\bar{\alpha}_a,\alpha^b] = \delta^b_a $. 
The latter may be interpreted as  creation and annihilation operators of the first string oscillator mode $(\alpha^a_{-1},\alpha^a_1)$ but this is not required. In particular, the index $a$ may run over an arbitrary set, including a single value. In order to preserve the  $\mathcal{N}=2$ worldline supersymmetry, we also introduce two complex fermionic variables $\psi^a,\bar{\psi}_a$ corresponding to the superpartners of the $\alpha$'s, with $(\psi^a)^{\dagger}=\bar{\psi}_a$ and  anticommutator, $\{\bar{\psi}_a , \psi^b \} = \delta^b_a$. 

The corresponding constraints in the absence of background fields can be obtained from the massless case in the last section requiring that the algebra of the constraints
\begin{subequations}
\begin{align}
    H = & \: -p^2 - m^2\label{h0} \alpha^a\bar{\alpha}_a - m^2 \psi^a \bar{\psi}_a\,, \\
    q = & \: \theta^{\mu}p_{\mu} + m\psi^a\bar{\alpha}_a\,, \label{q1}\\
    \bar{q} = & \:  \bar{\theta}_{\mu}p^{\mu} + m\alpha^a\bar{\psi}_a \,,\label{q2}
\end{align}
\end{subequations}
closes. 
They obey the same constraint algebra as for the massless case, that is
\begin{equation}
    \{q,\bar{q}\} = -H \,, \qquad 
\end{equation}
The resulting BRST charge\footnote{We denote the BRST charge with a 0 subscript to distinguish it from the one we will introduce in section \ref{sec:vec field}.}
\begin{equation} \label{eqn:BRST charge massive without vector}
    Q_0 = c H + \bar{\gamma} q + \gamma \bar{q} + \gamma \bar{\gamma} b \,,
\end{equation}
is thus nilpotent, Grassmann odd and has ghost number 1. Linear excitations of the dynamical system corresponding to $Q_0$ are elements of a linear representation space $V$, on which the BRST charge \eqref{eqn:BRST charge massive without vector} acts. We take this space to be spanned by  wave functions $\Phi(x^{\mu},c,\theta^{\mu},\gamma,\beta,\psi^a,\alpha^a)$ on which the operators $p_{\mu},b,\bar{\theta}_{\mu},\bar{\gamma},\bar{\beta},\bar{\psi}_a$ and $\bar{\alpha}_a$ act as derivatives. This makes $V$ into a Fock module with the vacuum $\ket{0}$ given by the constant function. 

We may again consider a reduced representation space by fixing the charge with respect to the  global $R$-symmetry current 
\begin{equation} \label{eqn: J current}
\begin{aligned}
    J & = \theta^{\mu}\bar{\theta}_{\mu} + \psi^a\bar{\psi}_a + \gamma\bar{\beta} - \bar{\gamma}\beta - \left( \frac{D}{2} - 1 \right) \\
    & = N_{\theta} + N_{\psi} + N_{\gamma} + N_{\beta} - \left( \frac{D}{2} - 1 \right) \,,
\end{aligned}
\end{equation}
which is conserved due to the $\mathcal{N}=2$ worldline supersymmetry. The last term in \eqref{eqn: J current} is a normal ordering constant. The vacuum $\ket{0}$ has charge $\left( 1 - \frac{D}{2} \right)$ under the current $J$, while the physical sector of the theory is given by the subspace of constant charge $\left( 2 - \frac{D}{2} \right)$. The reason for this will be clarified soon. 
A generic state in the subspace of charge $\left( 2 - \frac{D}{2} \right)$, linear in $\alpha$ or $\psi$ can be written as follows\footnote{At order zero in $\alpha$ and $\psi$ we recover the theory of massless vector fields described in \cite{Dai_2008}.} 
\begin{equation} \label{eqn: generic massive state}
\begin{aligned} 
    \ket{\Phi} = \big( & B_{\mu a}(x) \alpha^a \theta^{\mu} + \scalar_a(x) \psi^a + f^*_a(x)\alpha^a \gamma + g_a(x)\alpha^a \beta \\
    &+ B^*_{\mu a}(x)\alpha^a  c\theta^{\mu} + \scalar^*_a(x)c\psi^a + g_a^*(x)\alpha^a c\gamma + f_a(x)\alpha^a c\beta \big) \ket{0} \,.
\end{aligned}
\end{equation}
Unlike in string theory, this level-truncation is consistent since $Q_0$ preserves $N_\alpha+N_\psi$. We furthermore demand that the state $\ket{\Phi}$ is Grassmann odd and that it has ghost number 0, so that the first term in eq. \eqref{eqn: generic massive state} describes a set of vector potentials indexed by $a$. To understand what role each of the fields in eq. \eqref{eqn: generic massive state} plays, let us write down their Grassmann parity and ghost number.
\begin{center}
\begin{tabular}{ |c|c|c| } 
 \hline
 Field & Grassmann parity & Ghost number \\
 \hline
 $B_{\mu a}$ & $0$ & $0$ \\
 $\scalar_a$ & $0$ & $0$ \\
 $f_a$ & $0$ & $0$ \\ 
 $g_a$ & $1$ & $1$ \\
 $B_{\mu a}^*$ & $1$ & $-1$ \\
 $\scalar_a^*$ & $1$ & $-1$ \\
 $f_a^*$ & $1$ & $-1$ \\
 $g_a^*$ & $0$ & $-2$ \\
 \hline
\end{tabular}
\end{center}
\vspace{1ex}
So, $B_{\mu a}$ is a vector field, while $\scalar$ and $f$ are scalar fields. The field $g$ plays the role of a ghost. Fields denoted by a $*$ are the respective anti-fields of the BV-spectrum.

Physical states satisfy the condition
\begin{equation}
    Q_0 \ket{\Phi} = 0 \,.
\end{equation}
When imposing this on the generic state \eqref{eqn: generic massive state} we obtain a set of equations\footnote{Let us now denote $[\partial_{\mu},\Phi]$ simply by $\partial_{\mu} \Phi$ for all fields $\Phi$.}
\begin{subequations}
\begin{align}
    \Box B_{\mu a} - m^2 B_{\mu a} - i\partial_{\mu} f_a &= 0 \,,\\
    \Box \scalar_a - m^2 \scalar_a + mf_a &= 0 \,,\\
    f_a +i\partial^{\mu}B_{\mu a} - m\scalar_a &= 0 \label{eqn: auxiliaryMassive} \,,\\
    mg_a &= 0 \label{eqn: vanishing of the ghost} \,, \\
    \Box f^*_a - m^2 f^*_a + i\partial^{\mu} B^*_{\mu a} - m\scalar^*_a &= 0 \,.
\end{align}
\end{subequations}
Before moving on we observe that, in contrast to the massless case \cite{Bonezzi_2018}, the ghost field $g_a$ gets set to $0$. This highlights how by giving mass to the spin-$1$ particle we explicitly break gauge invariance. Equation \eqref{eqn: auxiliaryMassive} implies that the field $f$ is an auxiliary field since its equation of motion is algebraic. We can use the given equation to fix $f$ and obtain a set of equations for the remaining fields, that is,
\begin{subequations} \label{equations for real fields linear massive}
\begin{align}
    \partial^{\mu}(\partial_{\mu}B_{\nu a} - \partial_{\nu}B_{\mu a}) - m^2 B_{\nu a} &= im \partial_{\nu} \scalar_a \,,\\
    \Box\scalar_a &= im\partial^{\mu}B_{\mu a} \,.
\end{align}
\end{subequations}
Looking at eqs. \eqref{equations for real fields linear massive}, we observe that the real part of the vector field $B_{\mu a}$ couples to the imaginary part of the scalar field $\scalar_a$, and vice versa. If we set $\scalar_a$ to zero, we find that field $B_{\mu a}$ satisfies Proca equations. Note that so far we have not demanded for said field to be either real or Abelian. The upshot of this discussion is that at the linear level we can reproduce the theory for a massive vector field that can be both, complex and Lie algebra valued. As we will see in section \ref{sec:vec field}, this result does not hold beyond the linear level.

If we now let $\scalar_a$ be fully imaginary, and we substitute it with $-im\phi_a$, we find that eq. \eqref{equations for real fields linear massive} reproduce the Proca theory in its St\"uckelberg decomposition, where $\phi_a$ is the St\"uckelberg field. This implies that the theory at the linear level naturally contains the latter scalar as part of its spectrum. In the zero mass limit, we recover the BV-extended Maxwell theory together with a set of additional, decoupled scalar fields, $\varphi_a$.

Let us point out that as far as the linearised  BV-spectrum  and  equations are concerned, an identical system would be obtained removing $\alpha_a$ and $\bar\alpha_a$ in the supercharges in \eqref{q1}, \eqref{q2} and replacing the Hamiltonian constraint in \eqref{h0} by $-p^2+m^2$. This is akin of the construction in \cite{Bastianelli:2005uy} which amounts to a KK-reduction of the massless field in higher dimensions.

On a final note, let us observe that we could have gauged the symmetry generated by \eqref{eqn: J current}, also introducing a pair of ghosts for it, and including it into the BRST charge \cite{Bonezzi:2020jjq}. To select the correct sector of the Hilbert space we would have to introduce a Chern-Simons number, and this would select the correct sector throughout the time evolution of the states. This is particularly useful when computing scattering amplitudes, and it is the approach followed, for example, in ref. \cite{Bastianelli:2005vk, Bastianelli:2013pta}. However, when turning to a non linear analysis as we will do in sec. \ref{sec:vec field}, this not only makes computations highly cumbersome, but also does not avoid us from imposing the projection on the sector with the correct charge. Hence, we will avoid gauging the R-symmetry current, as we are rather interested in understanding the quantisation procedure and will not delve into the computation of scatterings of sort. The equivalence of the two descriptions can be inferred from the analysis given in \cite{Bonezzi:2020jjq}.

\section{Coupling to a background vector field}
\label{sec:vec field}
Let us now turn to a fully non-linear analysis of the equations of motion for the spacetime fields. To do this, we must introduce a vector field $B_{\mu}$ as a background for the worldline.\\
The vector field is a function of all bosonic coordinates, which are now $(x^{\mu}, \alpha_a , \bar{\alpha}^a )$. The simplest ansatz is a hermitian linear function of $\alpha$ and $\bar \alpha$, 
\begin{equation}\label{AnB}
    B_{\mu} = B_{\mu a} \alpha^a + B_{\mu}^{\dagger a} \bar{\alpha}_a \,.
\end{equation} 
The $\alpha$-dependence may be motivated by noting that if $B_\mu$ was independent of $\alpha$ and $\bar \alpha$, then the mass term in $H$ would not affect any dynamical equation since it commutes with the supercharges. On the other hand if $B$ is a non-linear function of $\alpha$, then $[H,q]$ will generate terms of even higher order in $\alpha$. One way to see this is to introduce a filtration, by dimension, on the operator algebra on $V$ as  in \cite{Grigoriev:2021bes} and to note that each power of $\alpha$ will increase the filtration by one. 

With the ansatz \eqref{AnB} the minimal coupling enters through the canonical momentum $ \Pi_\mu$ 
and casts the partial derivative to a covariant derivative:
\begin{equation}
    p_{\mu} \to \Pi_{\mu} = p_\mu  + B_{\mu a}\alpha^a + B^{\dagger a}_{\mu}\bar{\alpha}_a \,.
\end{equation}
Here $\dagger$ denotes hermitian conjugation. 

The minimal coupling to $B_\mu$ in $q$ and $\bar q$ then induces further non-minimal couplings in $H$. Concretely,
\begin{align} \label{eqn: full constraints}
    H = & \: -\Pi^{\mu}\Pi_{\mu} - m^2 \alpha^a\bar{\alpha}_a - m^2 \psi^a \bar{\psi}_a -2\kappa_1 \theta^{\mu} \bar{\theta}^{\nu} [\Pi_{\mu}, \Pi_{\nu} ] +2\kappa_2 m \left( \bar{\theta}^{\mu} B_{\mu a}\psi^a - \theta^{\mu}B^{\dagger a}_{\mu}\bar{\psi}_a \right) \,, \nonumber\\
    q = & \: \theta^{\mu}\Pi_{\mu} + m\psi^a\bar{\alpha}_a \,, \\
    \bar{q} = & \:  \bar{\theta}_{\mu}\Pi^{\mu} + m\alpha^a\bar{\psi}_a \nonumber\,,
\end{align}
where $\kappa_1$ and $\kappa_2$ will again be determined by nilpotency of 
\begin{equation} \label{eqn:full BRST charge massive}
    Q = c H + \bar{\gamma} q + \gamma \bar{q} + \gamma \bar{\gamma} b \,
\end{equation}
on a subspace $V_0$ of $V$ defined by the charge of the $R$-current $J$ in \eqref{eqn: J current}. Concretely we choose $V_0$ by setting the charge of $J$ to $2- \frac{D}{2}$ as in the massless case\footnote{In \cite{Dai_2008} the case with $D=4$ is considered, therefore only terms invariant under the $J$ current are kept.}. The physical interpretation of this choice is that we only allow for excitations as in \eqref{eqn: generic massive state}. This means that in $Q^2$ we need only consider terms with exactly one annihilation operator, namely in our convention barred operators and $b$. For $\kappa_1=\kappa_2=1$, the square of the BRST charge on this physical sector then reads
\begin{equation} \label{eqn: massive Q squared}
\begin{aligned}
     Q^2 \rvert_{J = 2-\frac{D}{2}} = \: & c\left[ H, \bar{\gamma}q + \gamma \bar{q} \right]\rvert_{J = 2-\frac{D}{2}} \\
     = \: & c \left( \bar{\gamma} \theta^{\nu} - \gamma \bar{\theta}^{\nu} \right) \left( - \left[ \Pi^{\mu}, \left[ \Pi_{\mu} , \Pi_{\nu} \right] \right] -m^2 B_{\nu a} \alpha^a - m^2 B^{\dagger a}_{\nu}\bar{\alpha}_a \right) \\
     & +m\bar{\gamma}\psi^a [\Pi^{\mu},B_{\mu a}]  -m\gamma \bar{\psi}_a [\Pi^{\mu}, B^{\dagger a}_{\mu}] \,.
\end{aligned}
\end{equation}
In the following subsections we analyse separately the case where $B_{\mu a}$ is an Abelian real field and the most general case where it is not.

\subsection{Abelian real vector field} \label{subsec: real abelian}
Let the field $B_{\mu a}$ now be real and Abelian, i.e.
\begin{equation}
    B_{\mu a} = B_{\mu a}^{\dagger} \,, \qquad [B_{\mu a}, B_{\nu b}] = 0 \,.
\end{equation}
Eqn. \eqref{eqn: massive Q squared} then simplifies considerably, 
\begin{equation} \label{eqn: abelian massive Q squared}
\begin{aligned}
    Q^2 \rvert_{J = 2-\frac{D}{2}} = \: & c \left( \bar{\gamma} \theta^{\nu} - \gamma \bar{\theta}^{\nu} \right) \left(  \partial^{\mu} (\partial_{\mu}B_{\nu a} - \partial_{\nu}B_{\mu a}) -m^2 B_{\nu a}\right) \left( \alpha^a + \bar{\alpha}^a \right) \\
    & -im\left(\bar{\gamma}\psi^a - \gamma \bar{\psi}^a\right) \partial^{\mu}B_{\mu a} \,.
\end{aligned}
\end{equation}
Since the first - and the second line in \eqref{eqn: abelian massive Q squared} have to vanish separately,  nilpotency of $Q$ imposes a set of two equations on the field $B_{\mu a}$ which reproduce the Proca theory of a massive vector boson
\begin{subequations} \label{eqn: abelian real Proca Theory}
\begin{align}
    \partial^{\mu}(\partial_{\mu}B_{\nu a}-\partial_{\nu}B_{\mu a}) - m^2B_{\nu a} &= 0 \label{proca equation real field} \,,\\
    m\,\partial^{\mu}B_{\mu a} &= 0 \,.\label{divergence free abelian proca}
\end{align}
\end{subequations}
The role of the bosonic oscillators $\alpha$ and $\bar{\alpha}$ is now clearer: they provide us with the correct coupling of the vector boson to the mass term in the Hamiltonian constraint. If we were to omit them, the mass term in the Hamiltonian would have no effect on the constraint algebra. As already mentioned in section \ref{Proca}, this is not an issue when analysing the theory at the linear level  \cite{Bastianelli:2005vk, Bastianelli:2005uy}. However, this approach is not applicable at the non-linear level, when analysing the nilpotency of the BRST charge.

\subsection{Non-Abelian and complexfied Proca field: a no-go} \label{subsec: non-Ab non real}
We now consider a generic field $B_{\mu a}$, which may be Lie algebra valued and possibly non-hermitian. Then, we must consider eq. \eqref{eqn: massive Q squared} in its full extent and ask for it to vanish. For Lie algebra valued $B_{\mu a} = B_{\mu a}^{\:i}\mathcal{T}^{\:i}$, the commutators then produce new terms of higher order in $\alpha_a$ and $\bar\alpha_a$. As the equation must hold order by order in $\alpha$ and $\bar{\alpha}$ separately, it will produce a whole set of terms which must vanish separately. However, terms of higher order in $\alpha$, i.e. $\mathcal{O}(\alpha^2,\alpha\bar{\alpha},\alpha^3,...)$, only vanish if the field $B_{\mu a}$ is Abelian. On top of that, the term of order zero in in $\alpha$ and $\bar{\alpha}$ only vanishes if the field $B_{\mu a}$ is hermitian. 

As a result, eq. \eqref{eqn: massive Q squared} can only vanish if the field $B_{\mu a}$ is real and Abelian, bringing us back to the case of sec. \ref{subsec: real abelian}. 
This result is consistent with what we already know from the usual formulation of quantum field theory: there is no extension of a non-Abelian gauge field with an explicit mass term which preserves the gauge symmetry. The present method provided us with a complementary way of showing how this happens. A non-Abelian background is inconsistent with BRST invariance of the worldline and thus propagation of one particle states. In fact, this conclusion is consistent with a related observation in \cite{Grigoriev:2021bes} (using an additional filtration) by noting that the the Proca field is in the same filtration as the graviton which is also required to be Abelian  (see \cite{Grigoriev:2021bes} for more details).

In closing this section we would like to point out that there is generalisation of the ansatz \eqref{AnB} of the form\footnote{See also \cite{Grigoriev:2021bes} for a related ansatz for a massless vector field.} $q=\theta^\mu\Pi_\mu+ m\psi^a\bar{\alpha}_a\;,\; \bar q =\bar{\theta}^\mu\bar \Pi_\mu + m\alpha^a\bar{\psi}_a$ with
\begin{align}
  \Pi_{\mu} &= p_\mu + B_{\mu a}\alpha^a + C_{\mu}\bar{\alpha}_a \,,\nonumber\\
  \bar \Pi_{\mu} &= p_\mu + \bar B_{\mu a}\alpha^a + \bar C_{\mu}\bar{\alpha}_a \,
\end{align}
where $B_{\mu a},\bar B_{\mu a}$ and similarly $C_{\mu a},\bar C_{\mu a}$ are independent and consequently $Q^\dagger\neq Q$. An explicit computation of $Q^2$ then shows that $B_{\mu a}=\bar B_{\mu a}=C_{\mu a}=\bar C_{\mu a}$. Thus, BRST-closure tells us that, the Proca field cannot be coupled to a dilaton. This in contrast to the massless Yang-Mills where such a coupling arises through  complexification by shifing the gauge potential, $A_{\mu} \to A_{\mu} - i\partial_{\mu}\varphi$ \cite{Grigoriev:2021bes,Bonezzi:2020jjq}.

\section{Coupling to a scalar} \label{section: coupling to a scalar}
As explained in \cite{Grigoriev:2021bes} the BRST quantisation procedure proves rather  useful to explore possible extensions of the worldline theory and see whether they are consistent with quantisation, i.e. if they are compatible with the nilpotency of the BRST charge. As an example, we will now consider the coupling of the theory to a set of scalar fields $\varphi^a$.

For this we consider a set of complex,  Abelian scalar fields $\varphi_a$ :
\begin{equation}
    (\varphi^a)^{\dagger} = \bar{\varphi}_a \,, \qquad [\varphi^a,\varphi^b] = 0 \,.
\end{equation}
As we discussed in sec. \ref{subsec: non-Ab non real}, we cannot couple a scalar field to $B_{\mu a}$ through a dilaton coupling. What we can do though, is couple it to $\psi^a$  which is formally equivalent to shifting the oscillator $\alpha_a$ by a spacetime-dependent commuting extension. That is, we modify the supercharges as follows:
\begin{subequations}
\begin{align}
    q = & \: \theta^{\mu}\Pi_{\mu} + m\psi^a\bar{\alpha}_a + m \psi^a\bar{\varphi}_a\,,\\
    \bar{q} = & \:  \bar{\theta}_{\mu}\Pi^{\mu} + m\alpha^a\bar{\psi}_a + m\bar{\psi}_a \varphi^a \,.
\end{align}
\end{subequations}
Again, we need to modify the Hamiltonian constraint too, to preserve nilpotency of $Q$ on $V_0$:
\begin{align} 
\begin{split}
    H = & \: -\Pi^{\mu}\Pi_{\mu} - m^2 \alpha^a\bar{\alpha}_a - m^2 \psi^a \bar{\psi}_a -m^2 \varphi^a\bar{\varphi}_a - m^2 (\varphi^a\bar{\alpha}_a + \bar{\varphi}_a\alpha^a) \\
    & - 2 \theta^{\mu} \bar{\theta}^{\nu} [\Pi_{\mu}, \Pi_{\nu} ] +2m \left( \bar{\theta}^{\mu} B_{\mu a}\psi^a - \theta^{\mu}B^{\dagger a}_{\mu}\bar{\psi}_a \right) \\
    & +2im \left( \theta^{\mu} \nabla_{\mu}\varphi^a\bar{\psi}_a + \bar{\theta}^{\mu}\nabla_{\mu}\bar{\varphi}_a \psi^a \right) \,.
\end{split}
\end{align}
Computing the commutator $[H,q]$ on $V_0$ we find\footnote{Looking at eq. \eqref{eqn: massive Q squared} we note that the only non trivial commutators when squaring $Q$ are $[H,q]$ and its hermitian conjugate. It is therefore sufficient to look at $[H,q]$ to check for the nilpotency of the BRST charge.}
\begin{equation} \label{presenceofstuckelberg}
\begin{aligned}
     {[H,q]} \rvert_{J = 2-\frac{D}{2}} = \: & \theta^{\nu} \big( -i \left[ \nabla^{\mu}, \left[ \nabla_{\mu} , \nabla_{\nu} \right] \right] -m^2 B_{\nu a} \alpha^a - m^2 B^{\dagger a}_{\nu}\bar{\alpha}_a \\
     & \qquad+im^2 \partial_{\nu} \varphi^a \bar{\alpha}_a -i m^2 \partial_{\nu}\bar{\varphi}_a\alpha^a \\
     & \qquad-im^2 B^{\dagger a}_{\nu}\bar{\varphi}_a - im^2B_{\nu a}\varphi^a + m^2 \varphi^a\partial_{\nu}\bar{\varphi}_a - m^2\partial_{\nu} \varphi^a\bar{\varphi}_a \big) \\
     & -im\psi^a [\nabla^{\mu}, B_{\mu a} + i[\nabla_{\mu},\bar{\varphi}_a]]  \,.
\end{aligned}
\end{equation}
The third line of eq. \eqref{presenceofstuckelberg} needs to  vanish in order to admit a non-trivial background for $\varphi$. For this to happen it turns out that the phases of fields $B$ and $\varphi$ must be spacetime-independent and they must differ by a factor of $\pi/2$. 
This means that, up to a global redefinition of the phase, the gauge field $B$ should be real and the scalar field $\varphi$ imaginary, or vice versa.

Let us explore the first possibility: $B_{\mu a}$ is real and $\varphi_a$ is  imaginary. We write $\varphi^a = i\phi^a$ and $\bar{\varphi}_a = -i\phi_a$. Equation \eqref{presenceofstuckelberg} then implies
\begin{subequations}
\begin{align}
    & \partial^{\mu}(\partial_{\mu} B_{\nu a}-\partial_{\nu}B_{\mu a}) - m^2B_{\nu a} -m^2\partial_{\nu}\phi_a = 0 \,,\\
    & \partial^{\mu} B_{\mu a} + \Box\phi_a = 0 \,.
\end{align}
\end{subequations}
Comparing this with \eqref{equations for real fields linear massive}  we see that this amounts to performing a St\"uckelberg decomposition of the massive gauge field, where $\phi_a$ is the St\"uckelberg field. This is in agreement with the linear analysis in  sec. \ref{Proca}. In the non linear analysis of this section, we can see how the St\"uckelberg field arises as  additional background field which, however, enters non-linearly in the equation \eqref{presenceofstuckelberg}.

\section{Spacetime action}\label{sec:action}
 We now aim to construct an action for $Q$ which reproduces the non-linear equations of motion on target space. In doing so, we seek to generalise the construction of \cite{Grigoriev:2021bes} for the massless case, where the bosonic oscillators $\alpha^a$ and $\bar \alpha^a$ as well as their superpartners $\psi^a$ and $\bar \psi^a$ are absent. For a non-Abelian vector field no satisfactory construction was found even in the massless case. Therefore, throughout this section, we will assume the BRST charge $Q$ to be as in section \ref{sec:vec field} with the extra condition that $B_{\mu a}$ is Abelian.
 
 In order to construct an action for $Q$ we need to define an inner product on the algebra of operators acting on $V$. As in \cite{Dai_2008,Grigoriev:2021bes} we utilise the  $(-1)$ ghost number state $\ket{-\beta} = -\beta\ket{0}\in V_0$ and then define a state $\ket{\Phi}\in V_0$ as
\begin{equation}
    \ket{\Phi} = \delta Q \ket{-\beta} = \left( \delta B_{\mu}\theta^{\mu} - \delta H c\beta \right) \ket{0}
\end{equation}
where $\delta Q$ and $\delta H$ are the deformations, due to the variation of the vector field, of the BRST charge and the Hamiltonian respectively. 
When comparing to the state in equation \eqref{eqn: generic massive state}, we notice that $-\delta H$ plays the role of $f$, the auxiliary scalar field. Therefore, when acting with $Q^{2}$ on the state $\ket{\Phi}$, we may expect to recover the full equations of motion up to ghosts and anti-fields, which do not appear here.
This suggests a cubic action as
\begin{equation}\label{sh}
    S = \bra{0}\Bar{\beta}QQQ\beta\ket{0} \,,
\end{equation}
where integration over spacetime coordinates and the $c$-ghost is understood. If the inner product above were cyclic and non-degenerate, the variation of this action would give the desired result. However, neither of these two properties are realised for our inner product.  Nevertheless, if we assume the gauge field $B_{\mu a}$ to be real we find
\begin{equation} \label{eqn: real Abelian Proca action}
    S =\bra{0}c\left[\left[H,\Bar{q}\right],q\right]\ket{0} = -\frac{1}{2}\int d^Dx \left( \frac{1}{2} B^a_{\mu\nu}B^{\mu\nu}_a + m^2 B_{\mu}^a B_a^{\mu} \right)\,,
\end{equation}
where $B^a_{\mu\nu} = \partial_{\mu}B_{\nu}^a - \partial_{\nu}B_{\mu}^a$ and we assumed the algebraic constraint $H=-\{q,\bar q\}$. 

Equation \eqref{eqn: real Abelian Proca action} reproduces the Proca action for a real field correctly. Next we would like to see what happens if we drop the reality condition on $B_{\mu a}$. The scalar product \eqref{sh} in the Hilbert space will project out all terms which do not have an equal number of $\alpha$'s and $\Bar{\alpha}$'s. Since $q$ acting on the ket in \eqref{eqn: real Abelian Proca action} produces an $\alpha^a$, terms independent of $\bar{\alpha}$ in $[H,\bar q]$ will be projected out in the scalar product and thus are not set to zero by the variational principle derived from the action $S$ in \eqref{sh}. As a consequence, in contrast to what we found in section \ref{subsec: non-Ab non real}, $B_{\mu a}$ need not be real. Indeed, for  $B_{\mu a}\neq  B_{\mu a}^\dagger$ we find instead of \eqref{eqn: real Abelian Proca action}, 
\begin{equation}
\begin{aligned}
    S = & \int d^Dx \left( - \partial^{\mu} (\partial_{\mu}B^{\dagger a}_{\nu} - \partial_{\nu}B^{\dagger a}_{\mu}) + m^2 B^{\dagger a}_{\nu} \right) B^{\nu}_a
    \\
    = & \int d^Dx \left( \frac{1}{2} B^{\dagger a}_{\mu\nu}B^{\mu\nu}_a + m^2B_{\mu}^{\dagger a}B_a^{\mu} \right)\,,
\end{aligned}
\end{equation}
which is a valid action for a massive, complex vector field but does not reproduce the reality constraint we found in section \ref{subsec: non-Ab non real}. This is a particular instance of the presymplectic formalism, discussed in \cite{Grigoriev:2021bes}, where a degenerate trace on the operator algebra eliminates some constraints  implied by nilpotency of $Q$.  

We note, in closing, that our construction of the action does not include the contribution of antifields and ghosts, and thus does not allow us to analyse the gauge symmetry breaking that we have observed at the linear level. An analysis which includes such contributions would require a modification of the BRST charge to include terms proportional to the antifields. However, such study goes beyond the scope of this work.

\section{Conclusions}\label{sec: conclusions}
In this work we have shown how coupling massive spin 1 particles to background vector fields and then requiring the nilpotency of the quantum BRST charge 
yields conditions for the background. It is of particular importance to understand the key point of these results: what is introduced as a purely non dynamical background, becomes dynamical as a requirement of consistency for the quantum theory. 

We first presented an analysis at the linear level, which led to expected equations, yet with no constraints neither on Abelianity nor on reality of the gauge field. On an interesting note, the St\"uckelberg field arises as a natural element of the spectrum of the linear theory.

When turning to the non-linear analysis of the massive particle, we draw inspiration from the results obtained in \cite{Dai_2008}. As in the massless case, we had to reduce the particle content to a subspace of the Hilbert space, identified by its charge under the global R-symmetry current. When coupled to a background vector field, conditions imposing Abelianity and reality of the vector field arise naturally as consistency requirements. We obtain equations \eqref{eqn: abelian real Proca Theory} reproducing Proca theory. It is worth mentioning that our analysis goes through as well if we replace the explicit mass term by a scalar field at the price of adding suitable extra non-minimal couplings in the Hamiltonian. 

We highlight that a system of equations appears which explicitly forbids a non-Abelian massive vector field, a result which matches our knowledge of massive Yang-Mills theory in the context of Lagrangian QFT.

Furthermore, we noticed that a coupling between the Proca field and the dilaton is explicitly forbidden, in contrast to the massless case \cite{Grigoriev:2021bes}. A different type of coupling to a scalar field was implemented in section \ref{section: coupling to a scalar}, leading to the appearance of a background St\"uckelberg field for the Proca field.

The strategy can be extended to fields of higher spin. This has been done for the $\mathcal{N}=4$ massless case coupled to a background spacetime metric \cite{Bonezzi_2018}. The quantisation yields the Einstein equations of motion, but there have been no attempts yet to extend this to massive gravitons. The same procedure for adding a mass implemented in this work could turn out to be successful also in the presence of larger extended supersymmetry.

Another interesting extension is to replace the mass term introduced here by a scalar field and to explore if Yukawa couplings can be reproduced in this way.

\acknowledgments
This work was funded by the Excellence Cluster Origins of the DFG under Germany’s Excellence Strategy EXC-2094 390783311. M.C. was supported by the German Academic Exchange Service (DAAD) under the funding programme "Study Scholarships for Graduates of All Discipline".


\bibliographystyle{unsrt}
\bibliography{literatur}



\end{document}